\documentstyle[preprint,aps]{revtex}
\begin{document}
\title{
ON SUPERFLUID PHASES OF COLD DECONFINED QCD
MATTER AT MODERATE BARYON DENSITY 
}
\author{
Ji\v{r}\'{\i} Ho\v{s}ek      \\
{\em Dept. Theoretical Physics, Nuclear Physics Institute } \\
{\em 25068 \v{R}e\v{z} (Prague), Czech Republic }\\
}
\maketitle
\baselineskip=14.5pt
\begin{abstract}
We present an overview of the arguments which lead to the picture
that at low temperatures the QCD matter not far above a critical
confinement-deconfinement density exists in one of several 
distinct superfluid phases exhibiting the quantum behavior 
on macroscopic scales.
\end{abstract}
\baselineskip=17pt
\newpage
\section{Introduction}
In QCD we trust\cite{wilczek1}. Con\-se\-quent\-ly, not far above a critical 
confinement-de\-con\-fi\-ne\-ment baryon density $n_c \sim 5n_{nucl.matter}$
and at low temperatures $T$ the deconfined QCD matter should be a rather
strongly interacting quantum many-colored-quark system. Its detailed actual
behavior in the considered region of the QCD phase diagram depends solely 
upon the details of the effective interactions relevant there.

It is natural to assume that the strong gluon interactions dress the tiny 
quark masses $m_u$ and $m_d$ (we restrict our discussion to the case of two
light flavors) into a common larger effective mass $m_*$, and become weak.
Residual interaction between the effective massive (quasi)quark excitations
$\psi^{a}_{\alpha A}$ ($a$-color, $\alpha$-Dirac, $A$-flavor $SU(2)$ indices)
can bona fide be described by appropriate short-range (approximately contact)
four-fermion interactions ${\cal L}_{int}$. Both $m_*$ and ${\cal L}_{int}$ are
to be fixed experimentally. Some parts of ${\cal L}_{int}$ have, however, a solid theoretical 
justification: (i) The instanton-mediated interaction of t'Hooft \cite{thooft}
\begin{equation}
{\cal L}^{I} = K_{I}\,[(\overline{\psi}\psi)^{2}+
(\overline{\psi}i\gamma_{5}\vec{\tau}\psi)^{2}-
(\overline{\psi}\vec{\tau}\psi)^{2}-(\overline{\psi}i\gamma_{5}\psi)^{2}]
\label{instanton}
\end{equation}
(ii) The Debye-screened chromoelectric one-gluon exchange interaction
\begin{equation}
{\cal L}^{D}= K_{D}\,
(\overline{\psi}\gamma_{0}\frac{1}{2}\lambda_{a}\psi)^{2}  \, ,
\label{chromoelectric}
\end{equation}
The resulting effective Lagrangian
\begin{equation}
 {\cal L}_{eff} = \overline \psi(i\gamma^{\mu}D_{\mu} - m_{*} + \mu \gamma_{0})\psi
 - \frac{1}{4}F_{a\mu\nu}F^{a\mu\nu} + {\cal L}_{int}
\label{effective}
\end{equation}
in which the gluon interactions are treated perturbatively (and neglected 
in the lowest approximation) defines a relativistic version of the Landau 
Fermi-liquid concept.

By assumption, ${\cal L}_{eff}$ is exactly $ SU(3)_c \times SU(2)_I \times U(1)_V
\times O(3)$ invariant. There is no approximate chiral $SU(2)$ symmetry of 
(\ref{effective}) which could be broken spontaneously. Hence, there should be no
light Nambu-Goldstone (NG) pions in dense and cold deconfined phase(s) of QCD.
It would be also misleading to think of $\psi$ and $m_*$ as of the constituent 
quark and of the constituent mass. There is nothing they might constitute.

At present, there are no experimental data, either real or the lattice ones 
which would check our assumption, although they are both heavily needed. For our
at best semi-quantitative considerations the assumption is not, however, essential.
An alternative picture is that in the cold deconfined QCD matter the $u,d$ quarks 
stay approximately massless at the Lagrangian level as they were in the confined phase.
Discussion of the superfluid phases presented below applies also to this case.
On top of that it is, however, mandatory to ask (and answer) how the (approximate)
chiral symmetry is realized in this case.

The cold and dense deconfined QCD matter should exist in the interiors of the 
neutron stars, and optimistically also in the early stages of the relativistic 
heavy-ion collisions studied experimentally with much effort at present.
In the following we will present arguments that the matter governed 
by (\ref{effective}) should exist below certain critical temperature $T_c$ of the order
of 100 MeV in some of several distinct superfluid phases exhibiting the quantum
behavior on macroscopic scales.

\section{Cooper instability}

When scaling the fermion momenta in dense, low-T quantum many-fermion systems, both 
non-relativistic\cite{polchinski} and relativistic\cite{evans}, towards 
the Fermi surface, all interactions but one become irrelevant. This implies that such systems
should behave thermodynamically as a corresponding noninteracting gas of fermions 
with the effective mass. For example, the specific heat of such systems should 
grow linearly with temperature. Such a behavior is indeed observed in the 
non-relativistic low-$T$ electron gas in metals, and in the liquid $^{3}He$. We
are not aware of any experimental data in the relativistic systems.

The four-fermion interaction attracting fermions with the opposite momenta at
the Fermi surface is the only exception: Even if arbitrarily small, it causes the
(Cooper) instability of the perturbative ground state with respect to spontaneous 
condensation of the fermion Cooper pairs with opposite momenta into a more
energetically favorable ground state. The new ground state, being by construction and by 
definition translationally invariant, has in the simplest 
case of the ordinary Bardeen-Cooper-Schrieffer (BCS) superconductor\cite{schrieffer}
the property
\begin{equation}  
\langle \psi^{+}_{\alpha}(x)(\sigma_{2})_{\alpha \beta} \psi^{+}_{\beta}(x) \rangle = \Delta \neq 0
\label{psipsi}
\end{equation}
It exhibits clearly the spontaneous breakdown of the $U(1)$ phase symmetry 
generated by the operator of the particle number, $N = \int d^3x \psi^{+}_{\alpha}\psi_{\alpha}$.

In the pre-BCS era the properties of superconductors were successfully 
described by beautiful phenomenological theories: (i) The Ginzburg-Landau (GL) theory\cite{schrieffer} 
associated to the ordered phase an order parameter - doubly 
charged ($e^* = 2e, m_{eff} = 2m_{*}$) complex 
scalar field developing a nonzero ground-state expectation value, $<\Phi> \neq 0$.
The supercurrent follows easily from the gauge-invariant kinetic term in the free energy
\begin{equation}
{\cal F}^{kin}_{GL} = \frac{1}{2m_{eff}}[(-i\vec{\nabla} + e^{*}\vec{A})\Phi ]^{+}(-i\vec{\nabla} + e^{*}\vec{A})\Phi
\end{equation}
A derivation of GL from BCS provided by Gor'kov\cite{schrieffer} implies in particular
$<\Phi> \sim \Delta$. (ii) Long before the GL theory F. London\cite{london} postulated
his famous equations (see Eqs.(\ref{london1}), (\ref{london2}) below) arguing as follows: "...   
the equations at which we have arrived are distinguished by their simplicity and symmetry 
in such a way that we could hardly avoid writing them down." Their form follows from
the GL theory by simple replacement $\Phi \rightarrow <\Phi >$. Lorentz-invariant 
quantum-field theory version of the GL theory is notoriously known as the Higgs (H) model.

The BCS ground state acts as the Fock's vacuum of the Bogolubov-Valatin (BV)\cite{schrieffer} 
true fermionic quasiparticles having the specific dispersion law
\begin{equation}
E_{\vec{k}} = \sqrt{\xi_{\vec{k}}^{2} + {\Delta}^2}
\label{bogo-vala}
\end{equation}
in which the energy gap $\Delta$ and $\xi_{\vec{k}}$ are self-consistently fixed by
the interaction. The quantities $\xi_{\vec{k}}$ are simply related to the perturbative 
form of the electron dispersion law ${\vec{k}}^{2}/2m$.
The (super)current $\vec{\jmath}_s$ acquires the form "$\vec{\jmath}_s = -(e^{*2}/m_{eff})|<\Phi>|^2 \vec{A}$" 
where $\vec{A}$ is the 
gauge potential, in contrast with the normal conductivity current $\vec{\jmath}_n = \sigma \vec{E}$
($\sigma$ is the conductivity). 
The quotation marks abbreviate a shorthand notation for the gauge-invariant 
London equations
\begin{equation}
\frac{\partial \vec{\jmath}_s}{\partial t} = (e^{*2}/m_{eff})|<\Phi>|^2 \vec{E}
\label{london1}
\end{equation}
\begin{equation}
rot \vec{\jmath}_s = -(e^{*2}/m_{eff})|<\Phi>|^2 \vec{B}
\label{london2}
\end{equation}
Below a critical temperature $T_c \sim 1K$ at which the spontaneous condensation 
of Cooper pairs sets in the quasifermion dispersion law (\ref{bogo-vala}) results 
in a abrupt change of the behavior of the specific heat on $T$ from linear to 
the exponential one. Another robust manifestation of superconductivity characterized
by the condensate $\Delta$ is the Meissner effect:  Magnetic field penetrates 
into the superconductor only within a London penetration 
length $\lambda_{L} = (m_{eff}/e^{*2}|<\Phi>|^2)^{1/2}$.
Operationally this phenomenon follows immediately from the Maxwell equation
$rot\vec{B} = \vec{\jmath}_s$ combined with Eq.(\ref{london2}): 
$({\nabla}^2 - \lambda_{L}^{-2})\vec{B} = 0$. The equation (\ref{london1})
accounts for perfect conductivity i.e., for stationary electric currents 
in absence of electric fields.

Analogous, but more sophisticated phenomena take place below $1mK$ in electrically neutral 
Landau Fermi liquid of $^3He$\cite{nobel3he}. We do not sharply distinguish between
superconductivity and superfluidity in many-fermion systems. The superconductor is
a superfluid having the perturbative long-range gauge interactions switched on.
Superfluidity in many-boson systems (for example in $^4He$) is, however, an entirely different story.

\section{Isotropic color-triplet superconductor}

In the cold and dense deconfined QCD matter governed by (\ref{effective}) the situation is 
physically very much the same as that in an electron Fermi liquid. The differences are 
rather technical: (i) Characteristic energies given by the chemical potential 
of the order of hundreds of $MeV$ require the relativistic description. (ii) The
quarks carry, besides spin, also the flavor and the color. Consequently, in 
comparison with (\ref{psipsi}) there are more possibilities for the local superfluid 
condensates in accord with the Pauli principle. (iii) The gauge fields are
both Abelian (photon) and non-Abelian (gluons). (iv) The origin of the effective 
interactions is different.

The most energetically favorable superfluid phase is for the realistic interactions ${\cal L}_{int}$ 
characterized by the condensate\cite{basics}
\begin{equation}
\langle \overline {\psi}_{\alpha aA}(x) {\epsilon}^{ab3} (\tau_2)_{AB} (\gamma_5C)_{\alpha \beta}
\overline {\psi}_{\beta bB}(x)\rangle = \Delta
\label{psibarpsibar}
\end{equation}
In a self-explanatory notation $C$ is the matrix of charge conjugation.
The condensate $\Delta$ corresponds to the ground-state expectation value $<\Phi^3>$ of a GLH order parameter
$\Phi^{c}(x)$ which transforms as a color triplet, isospin singlet, spin zero
complex field. Properties of this phase (and of its 3-flavor relative)  
are elaborated in the literature in most detail with several different interactions 
taken into account. Of interest are in particular: (i) form of the quasiquark 
dispersion law; (ii) properties of the NG excitations which must be present due to the fact
that the ground state in (\ref{psibarpsibar}) breaks the global $SU(3)_c$ down
to $SU(2)_c$; (iii) the fate of gluons if their interaction is switched on.

(i) It follows from (\ref{psibarpsibar}) that only the quarks of two colors
participate in Cooper pairing, while the third one  remains intact. Consequently, its
dispersion law remains ${\epsilon_{\vec{p\,}} }^2 = (| \vec{p\,} | \pm \mu)^2 + m_{*}^2$. 
The form of the dispersion law of the BV quasiquarks is
\begin{equation}
E(\vec{p\,})^2 = (\epsilon(\vec{p\,}) \pm \mu)^2 + |\Delta|^2
\label{relbv}
\end{equation}
where $\epsilon(\vec{p\,}) = \sqrt{{\vec{p\,}}^2 + m_{*}^2}$, and $\Delta$ is the 
energy gap fixed by the "gap equation"
\begin{equation}
\Delta = \Delta K \int \frac{d^3p}{(2\pi)^3}\left[ \frac{1}{\sqrt{(\epsilon(\vec{p\,}) +\mu)^2 + |\Delta|^2} } +
\frac{1}{\sqrt{(\epsilon(\vec{p\,})-\mu)^2 + |\Delta |^2  }}\right]  
\label{gapeq}
\end{equation}
In Eq.(\ref{gapeq}) $K$ is proportional to the coupling constant of the four-fermion 
interaction responsible for pairing. The UV divergence of the integral is eliminated 
by introducing a formfactor which mimics the asymptotic freedom. This phase is certainly 
interesting from the phenomenological point of view: In the alternative scenario
the instanton- mediated interaction (\ref{instanton}) gives rise simultaneously 
to the numerically acceptable spontaneous chiral symmetry breakdown, and to the 
phenomenologically interesting gaps $\Delta$ of the order of $100MeV$. The corresponding 
critical temperature is given by the universal BCS formula 
$\Delta/k_B T_c = \pi \exp(-\gamma)$,
where $\gamma \approx 0.5772$ is the Euler's constant.

(ii) It also follows from (\ref{psibarpsibar}) that this condensate breaks down 
spontaneously the $SU(3)_c \times U(1)_V$ global symmetry of the Lagrangian (\ref{effective})
down to $SU(2)_c$ (the gauge interactions switched off). According to the Goldstone theorem 
there must exist 5+1 gapless collective excitations in the spectrum. Clearly, they can 
only be excited by the quark bilinears. Their quantum numbers can be found with the help of the
Goldstone commutator (Q abbreviates the corresponding generators)
\begin{equation}
[Q, \overline \psi ... \overline \psi] = \overline \psi_a \epsilon^{ab3} \tau_2 \gamma_5C \overline \psi_b
\end{equation}
Not surprisingly, the NG quark bilinear combinations have the form ($A = 2,5,7$ ,
$\epsilon^{ab3} = i(\lambda_2)^{ab}$)
\begin{equation}
{\overline \psi\lambda_A \tau_2 \gamma_5C \overline \psi \pm H.c.}
\end{equation}.

(iii) If the gauge fields are switched on, and if they couple via the quasiquark loops 
to the NG bosons, the originally massless gauge bosons acquire masses by the 
Schwinger-Anderson mechanism. Corresponding calculations are rather involved, and
they were done only recently\cite{rischke}. In the GLH effective-field theory description 
of this color-superconducting phase the corresponding Higgs effect with a complex
color-triplet scalar field is elementary : The gauge fields $A_a^{\mu}$ stay massless for
$a=1,2,3$, and they acquire masses $m_a^2 = g^2|\Delta|^2$ and $m_a^2 = \frac{4}{3}g^2|\Delta|^2$
 for $a=5,6,7$ and $a=8$, respectively.

\section{The remaining three condensates}

Phase II is characterized by the condensate
\begin{equation}
\langle \overline {\psi}_{\alpha aA}(x)(\tau_3\tau_2)_{AB} (\gamma_5C)_{\alpha \beta}
\overline {\psi}_{\beta bB}(x)\rangle = \Delta_a \delta_{ab}
\label{phase2}
\end{equation}
As a previous one, also the condensate (\ref{phase2}) characterizes a relativistic isotropic 
superfluid: in both cases the quark bilinears look like the Lorentz scalars. Equation 
(\ref{phase2}) corresponds to the ground-state expectation value of a GLH order parameter 
$\Phi_{Iab}(x)$ which transforms as a color sextet, isospin triplet, spin zero complex field.

It is clearly possible to invent such an effective interaction which favors
namely this phase. In real QCD nobody knows. (This statement does not apply to the case 
of very high densities where the one-gluon exchange dominates, and its color-sextet 
channel is known to be repulsive.) To the best of our knowledge the details of this phase 
were not elaborated, although the generic form of its excitations it is easy to guess.
The phase is interesting mainly by spontaneous breakdown of the isospin symmetry. 
Since there is nobody who might "eat" them, the two corresponding gapless collective NG 
excitations remain in the physical spectrum, and become thermodynamically important.

Phase III is characterized by the condensate
\begin{equation}
\langle \overline {\psi}_{\alpha aA}(x)(\tau_2)_{AB}(\gamma_0\gamma_3C)_{\alpha \beta}
\overline{\psi}_{\beta bB}(x)\rangle = \Delta_{a}\delta_{ab}
\label{phase3}
\end{equation}
It corresponds to the ground-state expectation value of a GLH order parameter 
$\Phi_{ab\mu\nu}(x)$ which transforms as a color sextet, isospin singlet, antisymmetric
tensor field describing spin 1.

Phase IV is characterized by the condensate 
\begin{equation}
\langle \overline{\psi}_{\alpha aA}(x)(\tau_3 \tau_2)_{AB} \epsilon^{ab3}(\gamma_0\gamma_3C)_{\alpha \beta}
\overline {\psi}_{\beta bB}(x) \rangle = \Delta
\label{phase4}
\end{equation}
In its GLH description the order parameter $\Phi^a_{I\mu\nu}(x)$ transforms as a 
color triplet, isospin triplet, antisymmetric tensor field describing spin 1.
Remark: We speak of the order parameters as transforming as the Lorentz scalars or
antisymmetric tensors, respectively. It is, however, tacitly assumed that the derivative
terms in their corresponding GLH Lagrangians are only $O(3)$ invariant.

Phases III and IV are the anisotropic color superconductors, and they are most 
probably interesting primarily from the theoretical point of view. They exhibit
spontaneous breakdown of the rotational symmetry like ferromagnets. In relativistic
systems this is certainly not a very frequent phenomenon. It is possible only at finite density 
which itself breaks explicitly the Lorentz invariance. The less symmetry 
the more fragile the ordered phase is expected to be. Therefore, these phases should have
very low $T_c$. For an illustration we present just the form of the BV quasiquark dispersion laws (i = 1,2)
exhibiting nicely spontaneous breakdown of the rotational symmetry\cite{hosek}:
\begin{equation}
E_{i}^2(\vec{p\,}) = \epsilon^2(\vec{p\,}) + |\Delta|^2 + \mu^2 \pm 2\sqrt{\epsilon^2(\vec{p\,})\mu^2 
+ (p_1^2 + p_2^2 + m_{*}^2)|\Delta|^2}
\end{equation}

\section{Conclusions}

During the past two years so many interesting features of the color 
superconductors were pointed out and elaborated on that it is impossible to even 
mention them all. With apologies to those omitted in conclusion I consider appropriate
to discuss briefly two beautiful theoretical ideas:

(i) Color-flavor locking (CFL)\cite{cfl}. Variational analysis of a model with 
$SU(3)_c$ and exact global $SU(3)$ chiral symmetry reveals the following pattern 
of spontaneous symmetry breaking: There is a superfluid condensate corresponding to 
the lowest energy which 'locks' three groups $SU(3)$ i.e., c, L, R together
in such a way that only their diagonal $SU(3)$ subgroup (c + L + R ) remains 
unbroken. This is interesting by itself for there is a spontaneous chiral symmetry 
breakdown without $\langle \overline \psi \psi \rangle$ condensate. Properties 
of the low-energy excitations
in this deconfined phase are quite peculiar for they closely resemble hadrons
of the confined phase: (1) There are 8+1 baryons excited by the quasiquark fields, 
all having the chiral symmetry
breaking gap in their spectrum, and the integer charges. (2) There are 8+1 massive 
vector particles excited by the gauge fields, all having integer charges.
(3) There are 8+1 massless pseudoscalar collective excitations with integer charges.
(4) The electric charge corresponds to a particular linear combination of the 
diagonal $SU(3)$ generators in flavor and color spaces.

(ii) There is a new type of solution of the gap equation in the case of very high 
densities. In the standard case of moderate densities considered in Sect.3 the generic form
of the dependence of $\Delta$ on the effective coupling constant $K$ which is related to 
the gauge coupling constant $g$ as $K=g^2/{\Lambda}^2$ is
\begin{equation}
\Delta \sim \mu exp(- \frac{3\Lambda^2 \pi^2}{2 g^2 \mu^2})
\label{bcs}
\end{equation}
This is the characteristic BCS result.

As pointed out by Son\cite{son} and confirmed by others the case 
of very high densities is different. The exchange of unscreened magnetic 
gluons yields the dependence of $\Delta$ upon the small coupling constant $g$ in the form
\begin{equation}
\Delta \sim \mu g^{-5} exp(- \frac{3 \pi^2}{\sqrt2 g})
\end{equation}
This effect can provide numerically larger gaps than in previous estimates $\Delta \sim
100 MeV$ based on (\ref{bcs}).

It is gratifying to observe that our beloved QCD has a corner in its phase diagram
full of new phenomena associated with the existence of ordered quantum phases 
of the superfluid type. Moreover, there are good reasons to expect that
these phases can be described theoretically from the first principles. 

Let us assume optimistically that some sort of color superconductivity or superfluidity
does exist in the interiors of the neutron stars, and also in some events in
the relativistic heavy-ion collisions. Due to the macroscopic quantum nature 
of these phases it is perhaps justified to speculate that their experimental 
signatures might be brighter than those of the 'ordinary' quark-gluon plasma
above a superfluid $T_c$.

\section{Acknowledgements}
The author is grateful to the organizers for generous hospitality extended to him
in La Thuile.The work was supported by the grant GACR 202/0506.

\end{document}